# Normal heat diffusion in many-body system via thermal photons

Minggang Luo[1], Junming Zhao[1, 2, a)], Linhua Liu[3]

[1] *School of Energy Science and Engineering, Harbin Institute of Technology, Harbin 150001, China*

[2] *Key Laboratory of Aerospace Thermophysics, Ministry of Industry and Information Technology, Harbin 150001, China*

[3] *School of Energy and Power Engineering, Shandong University, Qingdao 266237, China*

**ABSTRACT**

A normal-diffusion theory for heat transfer in many-body systems via carriers of thermal photons is developed. The thermal conductivity tensor is rigorously derived from fluctuational electrodynamics as a coefficient of diffusion term for the first time. In addition, a convection-like heat transfer behavior is revealed in systems of asymmetric distribution of particles, indicating violation of Fourier's law for such system. Considering the central role of thermal conductivity in heat transfer, this work paves a way for understanding, analysis and manipulation of heat transfer in nanoparticle system via thermal photons with many-body interactions.

---

[a)] Corresponding author: jmzhao@hit.edu.cn (Junming Zhao)

Heat transfer via thermal photons at subwavelength separation distance (or near-field) can surpass by several orders of magnitude the heat flux exchanged between two blackbodies due to photon tunneling [1-4]. The tunneling of thermal photons enables many promising new application directions, such as near-field thermophotovoltaics [5], thermal logic circuits [6], heat-assisted magnetic recording [7], nanoscale temperature measurement [8] and nanoscale thermal management [9], etc. Near-field photonic heat transfer between two bodies were mostly considered in published works, both theoretically [10-14] and experimentally [15-19]. In system of many objects, however, heat transport will be significantly influenced by many-body interactions, which will suppress the exchanged heat flux [20-22] or even enable near-field energy transport to distance much larger than thermal wavelength [23, 24]. The picture of heat diffusion via near-field thermal photons in many body systems is vague. In a recent work, Ben-Abdallah *et al.* [25] discovered the super-diffusion behavior of heat transport in plasmonic nanostructure networks via near-field thermal photons, and a fractional-order diffusion equation was derived based on fluctuational electrodynamics. A naturally proposed question is: Is it possible to establish a normal-diffusion theory for near-field thermal photons from the first principle (fluctuational electrodynamics)?

Though the lack of a rigorous foundation of normal-diffusion theory, the kinetic theory for thermal photons were applied to calculate the photonic (radiative) thermal conductivity in many-body system [26-30], implying the validity of Fourier's law and the classic heat diffusion equation (HDE) for near-field thermal photons in many-body system. Two recent works [29, 30] analyzed the limitations of the kinetic theory by comparing to results of fluctuational electrodynamics, showed that the mismatch between the resonant mode and the thermally accessible modes limits the application of the kinetic theory. Alternatively, a formula to calculate the photonic thermal conductivity from many-body heat transfer theory was also proposed [31]. However, the validity and accuracy of the HDE for photonic heat transfer in many-body system has not been examined, especially for system with non-uniform distribution of particles.

In this letter, a normal-diffusion theory for heat transfer in many-body system is established rigorously based on fluctuational electrodynamics formulation. Both uniform and non-uniform distribution of particles are considered. A first-order term is discovered to appear in the governing equation for system with asymmetric distribution of particles, which reveals a convection-like heat transfer behavior of thermal photons. The photonic thermal conductivity tensor is derived naturally as the coefficient of the diffusion term.

To start the analysis, a many-body system in thermal non-equilibrium, composed of nanoparticles of different temperature exchanging energy only with thermal photons, is considered, which is depicted in Fig. 1(a). Our goal is to derive a normal-diffusion type governing equation of heat transport for the system (at continuum-scale, shown in Fig. 1(b)).

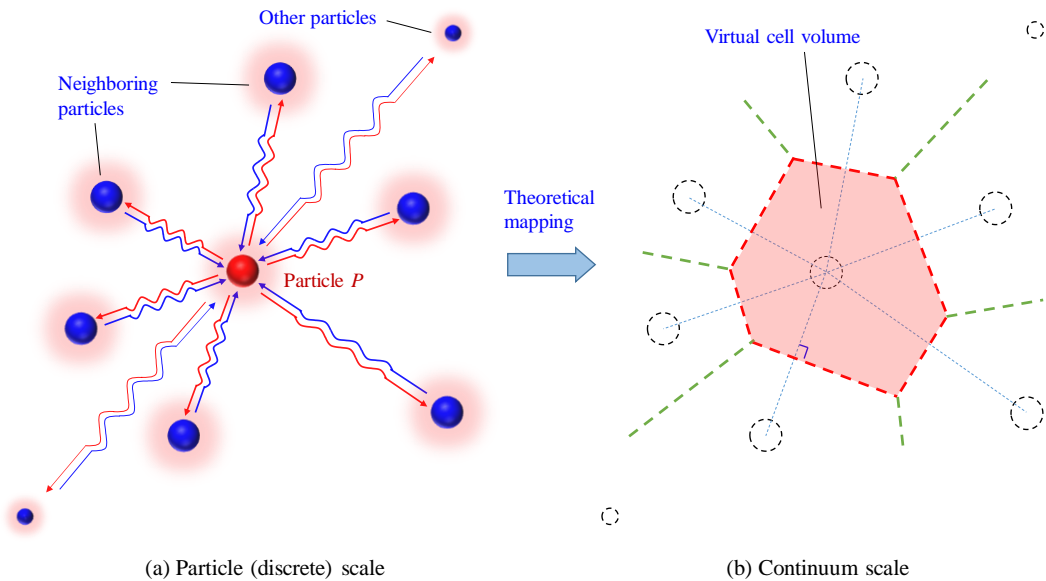

(a) Particle (discrete) scale  (b) Continuum scale

**FIG. 1** Schematic of near-field photonic heat transfer in nanoparticle system at **(a)** particle (discrete) scale and **(b)** continuum scale.

Considering only thermal photon mediated heat transfer, assuming the particles are small and temperature inside the particles are uniform, the energy balance of a selected particle ($P$) can be written as

$$\sum_{i=1}^{N_P} U_{i \leftrightarrow P}(T_i - T_P) + S_P = \frac{dE_P}{dt}, \quad (1)$$

where the subscript $i$ denotes the $i$-th particle (located at $\mathbf{r}_i$) exchanging heat with particle $P$ (located at $\mathbf{r}_P$,) as shown in Fig. 1(a), $N_P$ is the total number of particles in the system, $S_P$ [W] is the energy received from thermal bath and external source, $E_P = C_P T_P$ [J] is the internal energy of the particle, $C_P$ [J/K] is the particle heat capacity, $U_{i \leftrightarrow P}$ is the photonic thermal conductance between particle $i$ and $P$, defined as

$$U_{i \leftrightarrow P} = \frac{Q_{i \leftrightarrow P}}{T_i - T_P} = \frac{\varphi(T_i, l_i) - \varphi(T_P, l_i)}{T_i - T_P}, \quad (2)$$

where $Q_{i \leftrightarrow P}$ [W] is the net exchanged heat flux between $i$ and $P$, $l_i = |\mathbf{r}_i - \mathbf{r}_P|$ is the separation distance between particle $i$ and the particle $P$, $T_i$ and $T_P$ are the temperature of particle $i$ and particle $P$, respectively, $\varphi(l_i, T) = 3\int_0^{+\infty} \frac{d\omega}{2\pi} \mathcal{T}_{iP}(\omega) \Theta(\omega, T)$ is the transferred photonic heat flux by particle $i$ at $T$ to particle $P$, where $\Theta(\omega, T) = \hbar\omega / [\exp(\hbar\omega / k_B T) - 1]$ is the mean energy of Planck oscillator, $\mathcal{T}_{iP}(\omega)$ is the near-field transmission coefficient between particle $i$ and $P$, which can be calculated based on coupled dipole approach for small particles as [21, 32]

$$\begin{aligned}\mathcal{T}_{iP}(\omega) = \frac{4}{3}k^4 \Big[ &\mathrm{Im}(\chi_E^i)\mathrm{Im}(\chi_E^P)\mathrm{Tr}(G_{iP}^{EE} G_{iP}^{EE\dagger}) \\ &+ \mathrm{Im}(\chi_E^i)\mathrm{Im}(\chi_H^P)\mathrm{Tr}(G_{iP}^{EM} G_{iP}^{EM\dagger}) \\ &+ \mathrm{Im}(\chi_H^i)\mathrm{Im}(\chi_E^P)\mathrm{Tr}(G_{iP}^{ME} G_{iP}^{ME\dagger}) \\ &+ \mathrm{Im}(\chi_H^i)\mathrm{Im}(\chi_H^P)\mathrm{Tr}(G_{iP}^{MM} G_{iP}^{MM\dagger}) \Big]\end{aligned}, \quad (3)$$

considering both the contributions of electric and magnetic polarizations of particles, where $k$ is the wave vector, $\chi_E = \alpha_E - \dfrac{ik^3}{6\pi}|\alpha_E|^2$ and $\chi_H = \alpha_H - \dfrac{ik^3}{6\pi}|\alpha_H|^2$, $\alpha_E$ and $\alpha_H$ are the electric and magnetic polarizability (can be obtained from the extinction cross section with the help of the first order Lorenz-Mie coefficients [33, 34]), $G_{iP}^{EE}$, $G_{iP}^{ME}$, $G_{iP}^{EM}$ and $G_{iP}^{MM}$ are the Green's functions in many particles system. For particle made of dielectrics, the magnetic polarization is very weak and its contribution can be neglected, Eq. (3) can be simplified with considering only the first term [35]. While for metallic particles, both the contributions from electric and magnetic polarization should be considered [32].

Assuming the temperature distribution is locally continuous in the system, the temperature of the $i$-th particle ($T_i$) can be approximated from $T_P$ as

$$T_i = T_P + l_i \mathbf{e}_i \cdot \nabla T + \frac{1}{2}(l_i \mathbf{e}_i \cdot \nabla)^2 T + \frac{1}{6}(l_i \mathbf{e}_i \cdot \nabla)^3 T + O(l_i^3), \tag{4}$$

where $\mathbf{e}_i = (\mathbf{r}_i - \mathbf{r}_P)/l_i$ is a unit vector pointing from particle $P$ to particle $i$. According to Eqs. (2) and (4), the net exchanged heat flux between particles $i$ and $P$ can be calculated from

$$Q_{i \leftrightarrow P} = U_{i \leftrightarrow P}\left[l_i \mathbf{e}_i \cdot \nabla T + \frac{1}{2}(l_i \mathbf{e}_i \cdot \nabla)^2 T + \frac{1}{6}(l_i \mathbf{e}_i \cdot \nabla)^3 T + O(l_i^3)\right]. \tag{5}$$

Omitting the high order terms and substitute Eq. (5) into Eq. (1), the energy balance equation can be rewritten as

$$\sum_{i=1}^{N_P} U_{i \leftrightarrow P}\left[l_i \mathbf{e}_i \cdot \nabla T + \frac{1}{2}(l_i \mathbf{e}_i \cdot \nabla)^2 T\right] + S_P = \frac{dE_P}{dt}. \tag{6}$$

Consider a virtual cell with volume $V_{\text{cell}}$ occupied by the particle $P$ (sketched in Fig. 1(b)), which is location dependent for non-uniform distribution of particles, then divide Eq. (6) by $V_{\text{cell}}$ yields

$$\mathbf{h}_{\text{non-loc}} \cdot \nabla T + \mathbf{K}_{\text{non-loc}} : \nabla\nabla T + S = \frac{dE}{dt}, \tag{7}$$

where $\mathbf{h}_{\text{non-loc}} = \sum_{i=1}^{N_P} \frac{U_{i \leftrightarrow P}}{V_{\text{cell}}} l_i \mathbf{e}_i$ [W/(m²K)], $\mathbf{K}_{\text{non-loc}} = \sum_{i=1}^{N_P} \frac{U_{i \leftrightarrow P}}{2V_{\text{cell}}} l_i^2 \mathbf{e}_i \mathbf{e}_i$ [W/(mK)], $S = S_P/V_{\text{cell}}$ [W/m³], $E = E_P/V_{\text{cell}}$ [J/m³] are the continuum-scale physical parameters, the symbol $\mathbf{h}$ stands for a heat transfer coefficient vector accounting for asymmetric transfer process (called here *asymmetrical photonic heat transfer coefficient,* APHTC), $\mathbf{K}$ stands for the thermal conductivity tensor of thermal photons accounting for diffusion transfer process, $S$ is the volumetric heat source and $E$ is the energy density. The subscript '*non-loc*' (short for 'non-local') indicates the parameters are dependent not only on $T_P$, but also on temperature of all other particles $T_i$. It is noted that $U_{i \leftrightarrow P}$ is a function of $T_i$, $T_P$ and $l_i$.

In the following, the governing equation with ARHTC vector and photonic thermal conductivity tensor depending only on $T_P$ (local parameters) is derived. By using Taylor expansion of $\varphi(l_i, T_i)$ to the second order at $T_P$, $U_{i \leftrightarrow P}$ can be rewritten as

$$U_{i \leftrightarrow P}(T_i, T_P, l_i) = U_0(T_P, l_i) + \frac{1}{2} \frac{\partial U_0(T_P, l_i)}{\partial T_P}(T_i - T_P), \tag{8}$$

where $U_0(T_P, l_i) = \lim_{T_i \to T_P} U_{i \leftrightarrow P}(T_i, T_P, l_i) = \frac{\partial \varphi(T_P, l_i)}{\partial T_P}$. Substituting of Eq. (8) and $T_i - T_P \approx l_i \mathbf{e}_i \cdot \nabla T$ to Eq. (7) yields

$$\mathbf{h}'_{\text{loc}} \cdot \nabla T + \nabla T \cdot \left(\frac{\partial \mathbf{K}_{\text{loc}}}{\partial T}\right) \cdot \nabla T + \mathbf{K}_{\text{loc}} : \nabla \nabla T + S = \frac{dE}{dt}, \tag{9}$$

where the APHTC vector $\mathbf{h}'_{\text{loc}}$ and the photonic thermal conductivity tensor $\mathbf{K}_{\text{loc}}$ are defined respectively as

$$\mathbf{h}'_{\text{loc}}(T, \mathbf{r}) = \frac{1}{V_{\text{cell}}(\mathbf{r})} \sum_{i=1}^{N_P} U_0(T, l_i(\mathbf{r})) l_i(\mathbf{r}) \mathbf{e}_i, \tag{10}$$

$$\mathbf{K}_{\mathrm{loc}}(T,\mathbf{r}) = \frac{1}{2}\frac{1}{V_{\mathrm{cell}}(\mathbf{r})}\sum_{i=1}^{N_P} U_0(T,l_i(\mathbf{r}))l_i^2(\mathbf{r})\mathbf{e}_i\mathbf{e}_i. \tag{11}$$

Equation (9) is in non-divergence form, which can further be rewritten into a divergence form of normal-diffusion equation as (derivation details refer to the Supplemental Material)

$$\mathbf{h}_{\mathrm{loc}}\bullet\nabla T + \nabla\bullet(\mathbf{K}_{\mathrm{loc}}\bullet\nabla T) + S = \frac{\mathrm{d}E}{\mathrm{d}t}, \tag{12}$$

where the APHTC vector $\mathbf{h}_{\mathrm{loc}}$ is given as

$$\begin{aligned}\mathbf{h}_{\mathrm{loc}} =& \frac{1}{V_{\mathrm{cell}}(\mathbf{r})}\sum_{i=1}^{N_P} U_0(T,l_i(\mathbf{r}))\, l_i\mathbf{e}_i - \frac{1}{2}\frac{1}{V_{\mathrm{cell}}(\mathbf{r})}\sum_{i=1}^{N_P} l_i^2\frac{\partial U_0(T,l_i(\mathbf{r}))}{\partial l_i}\nabla l_i\bullet\mathbf{e}_i\mathbf{e}_i \\ &- \frac{1}{V_{\mathrm{cell}}(\mathbf{r})}\sum_{i=1}^{N_P} U_0(T,l_i(\mathbf{r}))l_i(\mathbf{r})\nabla l_i\bullet\mathbf{e}_i\mathbf{e}_i - \frac{1}{2}\nabla\left(\frac{1}{V_{\mathrm{cell}}(\mathbf{r})}\right)\bullet\sum_{i=1}^{N_P} U_0(T,l_i(\mathbf{r}))l_i^2(\mathbf{r})\mathbf{e}_i\mathbf{e}_i\end{aligned}. \tag{13}$$

Note that the obtained governing equation (Eq. (12) and Eq. (9)) are the core findings of the present study, which are the derived norm-diffusion heat transfer equation for many-body systems with many-body APHTC vector and photonic thermal conductivity tensor given by Eqs. (10), (13), and (11), respectively. The above derivation is general and can be applied to one-, two- and three-dimensional (1D, 2D and 3D) system of nanoparticles.

The obtained divergence-form normal-diffusion equation (Eq. (12)) is in mathematical form similar to the HDE. However, an obvious difference, namely, an additional convection-term (the first-order term) appears. Since the APHTC vector $\mathbf{h}_{\mathrm{loc}}$ is zero for uniformly (or symmetrically) distributed particulate system, the convection-term accounts only for the asymmetric heat transfer. The derivation indicates that heat transfer in many-body system via thermal photons not only contains a diffusion process, but also a convection process. The latter shows the heat transfer has a preferential direction, caused by the asymmetric distribution of particles in the system. From the derivation, the thermal conductivity tensor of thermal photons ($\mathbf{K}_{\mathrm{loc}}$) is revealed naturally as the coefficient of the diffusion term, and which is the only parameter characterize the heat transfer process for uniform (isotropic) distributed particulate systems (where $\mathbf{h}_{loc} = 0$), demonstrating the validity of Fourier's law for the diffusion process. However, if $\mathbf{h}_{loc} \neq 0$, only the thermal conductivity $\mathbf{K}_{loc}$ is insufficient to characterize both the diffusion and convection-type heat transfer processes, indicating the violation of the Fourier's law.

In the following, we apply the developed normal-diffusion theory to analyze the steady-state heat transfer in 1D, 2D and 3D nanoparticle networks in vacuum. The calculated temperature field distribution based on the normal-diffusion theory are verified by comparison with the results solved directly from the exact fluctuational electrodynamics formulation [32, 35]. The calculated photonic thermal conductivity from the new theory [$\mathbf{K}_{loc}$, Eq. (11)] are also compared with that predicted using kinetic theory. During the calculation, the separation distance between any two particles in the systems is limited no shorter than $3a$ to ensure the validity of the dipole approximation [36].

Firstly, photonic heat transfer in one-dimensional linear chains of uniform distribution of particles are considered. The left and right end of the chains are connected to hot (at 400 K) and cold reservoir (at 300 K), respectively, as shown in the inset of Fig. 2(a). The chains are composed of 60 nanoparticles, and three kinds of materials, namely, SiC, gold, and hBN, are considered. Structure information of the particle chains and optical properties of the materials considered in this work are consistent with that of the Refs. [29, 30]. The radius of nanoparticle ($a$) are 5 nm and 25 nm, and the lattice spacing $s$ are $4a$ and $7a$ for gold and hBN chain, respectively. While for SiC chain, $a$ is 25 nm and $s$ is $3a$. A heating power is added to each particle in the chains, which are (at particle-scale, $S_P$) $3\times10^{-13}$ W, 0 W and $2\times10^{-15}$ W for SiC, gold, and hBN nanoparticles, respectively. The equivalent volumetric heat source (at continuum-scale, $S$) is calculated as $S = S_P/V_{cell} = S_P/\pi s a^2$ by definition. The temperature distributions in the nanoparticle chains solved based on the developed normal-diffusion theory [Eqs. (12) and (11)], and the HDE with thermal conductivity calculated using the kinetic theory, are shown in Fig. 2 (a).

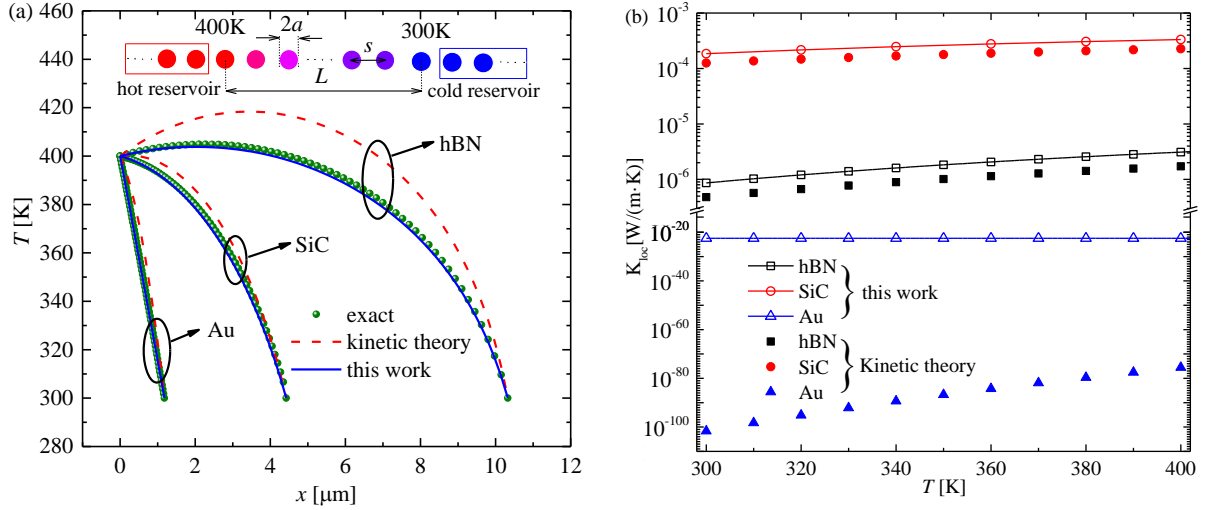

**FIG. 2 (a)** Temperature distributions along the nanoparticle chains solved based on the normal-diffusion theory and the classical heat diffusion equation with the thermal conductivity calculated using kinetic theory. The results calculated from fluctuational electrodynamics at particle-scale are shown as 'exact'. The left and right end of the chains are connected to hot (at 400 K) and cold reservoir (at 300 K). The chains contain 60 nanoparticles, which are composed of SiC ($a$=25nm, $s$=3$a$) [30], Au ($a$=5nm, $s$=4$a$) and hBN ($a$=25nm, $s$=7$a$) [29], and the nanoparticles are heated with power $S_P$ = $3\times10^{-13}$ W, 0 W and $2\times10^{-15}$ W, respectively. **(b)** Photonic thermal conductivity as a function of temperature $T$ for different nanoparticles chains, calculated from the normal-diffusion theory and the kinetic theory.

Results show that the temperature distribution obtained from the proposed normal-diffusion theory agrees very well with the exact result, which gives the maximum relative errors of 0.8%, 0.4% and 0.5% for SiC, Au and hBN chains, respectively. Dependence of the photonic thermal conductivity on temperature for the three chains are shown in Fig. 2 (b). The photonic thermal conductivity for SiC and hBN chains are much larger than that of the Au chain, ascribed to the strong thermally accessible coupling in the former, as compared to the latter. The kinetic theory is completely inappropriate for calculating the photonic thermal conductivity of the Au chain due to the mismatch between the resonance frequency and the Planck's window [29]. Generally, HDE combined with kinetic theory overestimates the temperature, as shown in Fig. 2 (a), which is attributed to the inaccuracy of predicted photonic thermal conductivity (shown in Fig. 2 (b)). Details on the solution of the normal-diffusion equation in a linear chain are provided in the Supplemental Material.

Then, photonic heat transfer in systems of non-uniform distributions of nanoparticles are considered, to demonstrate the effect of the convection-type heat transfer process of near-field thermal photons. The material of nanoparticles is hBN. The particles are packed closely on the left side (at 400 K) and sparsely towards the right side (at 300 K). Two chains with different non-uniformity (Case 1 and Case 2), and a uniform-distributed chain (Case 3), are considered. The temperature distributions in the three nanoparticle chains calculated based on the normal-diffusion theory [Eq. (9)], and the HDE combined with $\mathbf{K}_{\text{loc}}$, are shown in Fig. 3 (a). Note that the HDE can be considered the omission of the convection term in the normal-diffusion equation [Eq. (12)], hence it cannot well account for the convection-type heat transfer in systems of non-uniform distributions of nanoparticles. For Case 3, the particles are uniformly distributed, the HDE predict the temperature distribution with good accuracy. While for Case 1 and Case 2, the particles are distributed unevenly, the HDE significantly underestimates the temperature distributions. In all the cases, the normal-diffusion theory shows good accuracy compared to the exact results. Strong non-linearity of temperature distribution is observed for the chain with high asymmetry of particle distribution. At particle-scale, the temperature of a nanoparticle in both Case 1 and Case 2 is more significantly affected by its left neighbor (where the particles are denser) than the right neighbor nanoparticles, since the near-field interaction from its left neighbor is much stronger than that from the right. Hence a preferred direction heat transfer exists along the chain from left to right, which is a convection-like process with convection direction to the right. The HDE has not taken this convection-like process into consideration, which is thus not sufficient to describe the photonic heat transfer in asymmetrical systems, and the APHTC vector $\mathbf{h}_{loc}$ has to be included to fully characterizes the heat transfer in many-body system.

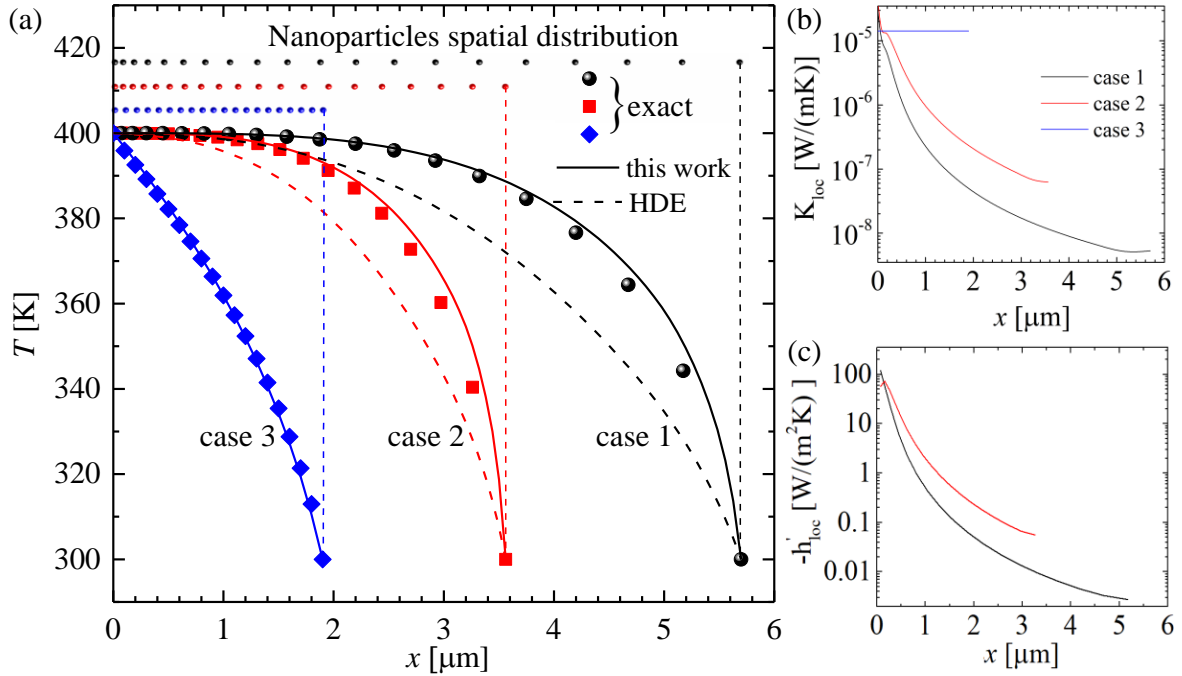

**FIG. 3** (a) Temperature distributions in the three hBN nanoparticle chains, the leftmost particle is fixed at 400 K, the rightmost particle is fixed at 300 K, which are calculated based on the normal-diffusion theory [Eq. (9)], and the classical heat diffusion equation (HDE). The results from exact fluctuational electrodynamics are shown as reference. The three different particle distributions are shown as inset. The radius of nanoparticle is 25 nm. The number of particles in the chain is 20. The particle coordinates are provided in the Supplemental Material. (b) The spatial dependent photonic thermal conductivity $K_{loc}$ and (c) the APHTC $\mathbf{h'}_{loc}$, at 300 K.

The photonic thermal conductivity and the APHTC vary significantly with location, as shown in Fig. 3 (b) and (c) (details on calculation refer to the Supplemental Material). The negative value of ARHTC shows a direction of convection-like heat transfer from left to right. Both the photonic thermal conductivity and the ARHTC decrease dramatically (several orders of magnitudes) from left side to the right side of the chain, consistent with the decreasing of near-field interaction with particle separation distance. From the above observation, special attention should be paid on near-field photonic heat transfer in non-uniform distributed particulate systems, where the HDE may not work well.

Finally, the normal-diffusion theory is applied to 2D and 3D nanoparticle systems, to demonstrate the derived Eq. (11) is a powerful tool for exploring the thermal conductivity of thermal photons in complex multidimensional nanoparticle system. For 2D system, four different structures, namely, random, square, hexagonal and honeycomb lattice distributions are considered (as shown in insets of Fig. 4). For 3D system, the nanoparticles are randomly distributed. Based on the proposed normal-diffusion theory of thermal photons, thermal conductivity is the only parameter to characterize the heat transfer in (statistically) uniformly distributed particulate systems, since $\mathbf{h}_{\text{loc}} = 0$. Photon tunneling is very sensitive to separation distance, hence average separation distance between neighboring nanoparticles, $c$, is a key parameter that affects the heat transfer process. An alternative parameter is the volumetric filling fraction $f_v$. The thermal conductivity of the considered nanoparticle systems as a function of $f_v$ and $c$ are shown in Fig. 4. The material of the nanoparticles is SiC and the radius is 100 nm. The photonic thermal conductivity is evaluated at system temperature 300 K. The photonic thermal conductivity of 1D linear chains is also shown for reference. From computational resources consideration, the calculations are performed for systems of the size of 1000 nanoparticles.

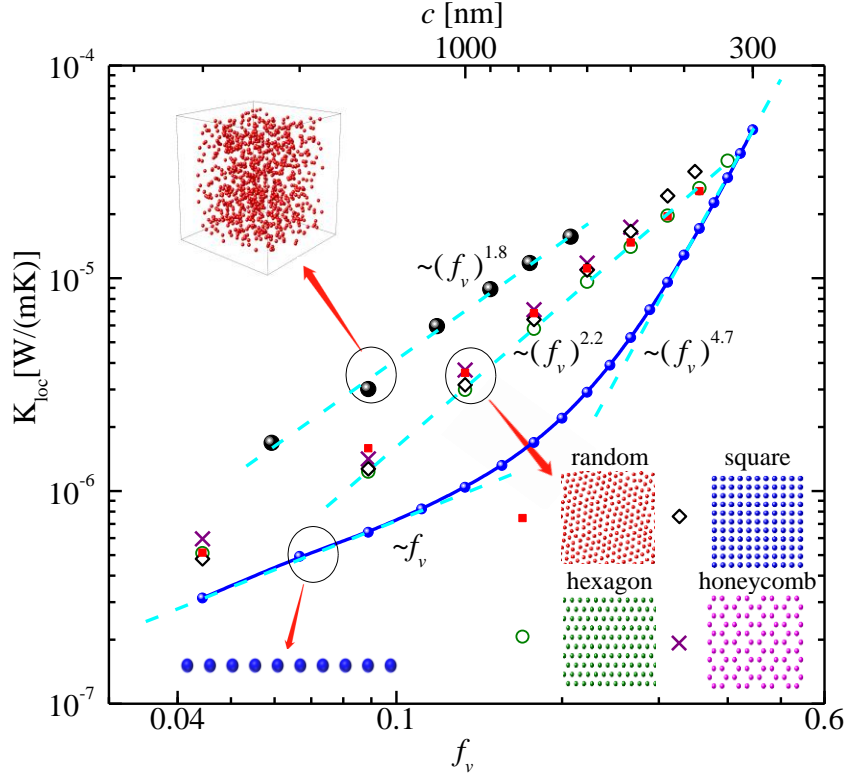

**FIG. 4** Thermal conductivity of thermal photons as a function of filling fraction for 1D, 2D and 3D nanoparticle system (evaluated at 300 K). The material of the nanoparticles is SiC and the radius is 100 nm. Diagrams of the structures of the considered nanoparticle systems are given in the insets. Four different structures (random, square, hexagonal and honeycomb lattice distributions) are considered for the 2D systems. For the 3D system, a random distribution is considered. The separation distance $c$ of particles with $f_v$ for the 1D case is added as the top axis. The calculations are performed for systems of the size of 1000 nanoparticles. Analysis on the spectral radiative thermal conductivity is provided in the Supplemental Material for reference.

As shown in Fig. 4, for the same $f_v$, the photonic thermal conductivity for 3D nanoparticle system is significantly greater than that of the 2D system and that for the 2D systems is further greater than the 1D chains. This can be explained that more neighboring particles existing in the high-dimensional nanoparticle system provides more heat transfer channels, which is also consistent with the reported observation in the systems of only a few nanoparticles [35] or hundreds of nanoparticles [22]. For 1D linear chains, photonic thermal conductivity varies radically with the filling fraction $f_v$, from $\sim f_v$ to $\sim f_v^{4.7}$ with increasing $f_v$. The $f_v$ dependence of photonic thermal conductivity for the 2D and 3D system are observed to be different from that of the 1D system. It shows no abrupt change of the trend of $f_v$ dependence of photonic thermal conductivity, which may be due to limited parameter range. When considering sufficiently sparse 2D and 3D system where near-field interaction is weakened, the $f_v$ dependence of photonic thermal conductivity is expected to have a radical change as compared to that of the dense systems. For 2D system, the photonic thermal conductivity and its dependence on $f_v$ for different lattice structures is different. For high $f_v$ (~0.3), the photonic thermal conductivity of square lattice shows the highest value, whereas that of the random distribution shows lowest value. For low $f_v$ (~0.045), that of honeycomb lattice shows the highest value, while that of square lattice shows lowest value. The results illustrate the possibility of designing the photonic thermal conductivity by structure optimization.

In summary, a normal-diffusion theory for heat transfer in nanoparticle systems via thermal photons is developed. Interestingly, a first-order term appears in the normal-diffusion equation for the system with asymmetrical distribution of particles, which reveals a convection-like heat transfer behavior, indicating violation of Fourier's law for such system. The photonic thermal conductivity tensor is derived as the coefficient of diffusion term and a new formula (Eq. (11)) for calculation is obtained. For uniformly distributed nanoparticle systems, the normal-diffusion equation reduces to the HDE. The accuracy of the proposed theory is verified based on the fluctuational electrodynamics solution at particle scale. The temperature distribution predicted by the proposed theory agrees very well the exact solutions both for systems of uniform and non-uniform distribution of particles. Whereas, the HDE cannot well capture the convection-like heat transfer behavior of near-field thermal photons in systems of non-uniformly distributed nanoparticles. The derived Eq. (11) is a powerful tool for exploring the thermal conductivity of thermal photons in multi-dimensional particulate systems. In prospect, the characteristics of photonic thermal conductivity in nanoparticle system of variety of materials, structures, and dimensions, especially the non-uniform distributed system, the effect of many-body interaction, and coupled-mode heat transfer analysis with considering near-field photonic carrier, etc., can be explored based on the proposed theory.

The support of this work by the National Natural Science Foundation of China (No. 51976045) is gratefully acknowledged.